\pdfoutput=1
\documentclass{JINST}

\title{Imaging with SiPMs in noble-gas detectors}
\author{N.~Yahlali$^{a}$\thanks{Corresponding author}, 
L.M.P.~Fernandes$^{b}$, 
K.~Gonz\'alez$^{a}$,  
A.N.C. Garcia$^{b}$,
A.~Soriano$^{c}$

(On behalf of the NEXT collaboration)
\\
\llap{$^{a}$}
Instituto de F\'isica Corpuscular (IFIC), CSIC \& Universitat de Val\`encia\\
Calle Catedr\'atico Jos\'e Beltr\'an, 2, 46980 Paterna, Valencia, Spain\\
\llap{$^{b}$}
Departamento de Fisica, Universidade de Coimbra\\
Rua Larga, 3004-516 Coimbra, Portugal\\
\llap{$^{c}$} Instituto de Ciencia Molecular (ICMOL) \\
Catedrático José Beltran 2, 46980 Valencia, Spain.\\

E-mail:  \email{nadia.yahlali@ific.uv.es}
}

\abstract{Silicon photomultipliers (SiPMs) are photosensors widely used for imaging in a variety of high energy and nuclear physics experiments. In noble-gas detectors for double-beta decay and dark matter experiments, SiPMs are attractive photosensors for imaging. However they are insensitive to the VUV scintillation emitted by the noble gases (xenon and argon).  This difficulty is overcome in the NEXT experiment by coating the SiPMs with tetraphenyl butadiene (TPB) to convert the VUV light into visible light. TPB requires stringent storage and operational conditions to prevent its degradation by environmental agents. The development of UV sensitive SiPMs is thus of utmost interest for experiments using electroluminescence of noble-gas detectors. It is in particular an important issue for a robust and background free $\beta\beta0\nu$ experiment with xenon gas aimed by NEXT.
The photon detection efficiency (PDE) of UV-enhanced SiPMs provided by Hamamatsu was determined for light in the range 250\textendash500 nm. The PDE of standard SiPMs of the same model (S10362-33-50C), coated and non-coated with TPB, was also determined for comparison. 
In the UV range 250\textendash350~nm, the PDE of the standard SiPM is shown to decrease strongly, down to about 
3\%. 
The UV-enhanced SiPM without window is shown to have the maximum PDE of 44\% at 325 nm and 30\% at 250 nm. The PDE of the UV-enhanced SiPM with silicon resin window has a similar trend in the UV range, although it is about 30\% lower.
The TPB-coated SiPM has shown to have about 6 times higher PDE than the non-coated SiPM in the range 250\textendash315 nm. This is however below the performance of the UV-enhanced prototypes in the same wavelength range. 
Imaging in noble-gas detectors using UV-enhanced SiPMs is discussed.}


\begin{document}
%\linenumbers

%%%%%%%%%%%%%%%%%%%%%%%%%%%%%%% %%%%%%%%%%%%%%%%%%%%%%%%%%%%%%%%%%%%%%%%%%                                      Section 1                                    %%%%%%%%%%%%%%%%%%%
%%%%%%%%%%%%%%%%%%%%%%%%%% %%%%%%%%%%%%%%%%%%%%%%%%%%%%%%%%%%

\section{Introduction}  \label{sec:intro}

Silicon Photomultipliers (SiPMs) coupled to scintillators are nowadays widely used for imaging in a variety of high energy and nuclear physics experiments \cite{{Garutti:2011},{Ieki:2009}}. 
These devices offer very attractive features for optical imaging in large-scale electroluminescent noble-gas time projection chambers (TPCs) used in double-beta decay or dark matter searches \cite{{Alvarez:2012-2},{McConkey:2011}}, mainly cost-effectiveness, a response level comparable to that of photomultiplier tubes (PMTs) and radiopurity. 
SiPMs are however insensitive to the VUV scintillation of noble gases (peak at 175 nm in xenon and at 129 nm in argon), which requires the use of a wavelength shifter (WLS) to convert the VUV light into visible light to which these devices are most sensitive (peak at 430 nm). Tetraphenyl butadiene (TPB) is used as a WLS, deposited on the SiPMs by vacuum evaporation 
\cite{Alvarez:2012-1}, or dissolved in a plastic binder and painted on the active surface of the photosensors
 \cite{Lightfoot:2009}.  This solution is proved to ensure an adequate response of the SiPMs to the electroluminescence light of xenon and argon gases \cite{McConkey:2011}\textendash\cite{Lightfoot:2009}. The TPB requires on the other hand, stringent storage and operational conditions to avoid ageing due to hydration and oxidation and to ensure the long-term stability of the coatings. Hence, UV sensitive SiPMs would be extremely attractive. 

The spectral dependence of the photon detection efficiency (PDE) of SiPMs is mainly driven by the photon absorption length in silicon which varies from 10 nm to a few microns for wavelengths between 300 nm and 700 nm and is less than 10 nm for shorter wavelengths \cite{{Henke:data},{LeiShi:2012}}. 
When the photon attenuation length is below the micron, a high sensitivity of a silicon-based photodetector can only be achieved if the photo-detection region is close to the surface.
The research efforts for the development of UV-sensitive silicon-based photodetectors are directed towards increasing the sensitivity near the silicon surface by creating damage-free ultra-shallow junctions (< 200 nm). The possibility of an additional electric field close to the detector surface, which transports the photo-generated charges to the depletion region allowing their collection, is also considered \cite{{LeiShi:2012},{LeiShi:2011}}.
 
We present a study of the PDE of UV-enhanced SiPM prototypes provided by Hamamatsu, also known as multi-pixel photon counters (MPPCs).
The SiPMs considered are different versions of the commercial MPPC S10362-33-050C from Hamamatsu, whose
main characteristics are an active area of 3~mm$\times$3~mm, with 3600 pixels of 
50~$\mu$m$\times$50~$\mu$m each. The fill factor is 61.5\% and the typical voltage range for operating the MPPC is 70$\pm$2 V.  
The standard MPPC, so-called STD in the present work, has a window made of an organic polymer which protects the silicon layer
from possible damages derived from environmental agents as dust and moisture. It has however the disadvantage of absorbing UV photons. 
The so-called STD-TPB sample is a standard MPPC coated with 0.1 mg/cm$^2$ ($130\pm10$~nm) of TPB, deposited by vacuum-evaporation following the protocol described in \cite{Alvarez:2012-1}.      
The so-called UVE-SPL sample is a version of the standard MPPC with a UV-enhanced silicon layer and
without a protective window.
 The fourth sample tested, so-called UVE-SIRESIN, is provided with a UV-enhanced silicon layer and a silicon resin window. 
Silicon resin consists of highly branched silicon structures combined with organic polymers \cite{Vanlathem:2006}, which have numerous outstanding properties, e.g. hydrophobic properties, high thermo-mechanical stability and high UV transparency compared to organic polymers \cite{DeGroot:2004}.

 In the following sections the PDE measurement method used is described. Then the PDE measurements in the spectral range 250\textendash500 nm are presented. The use of the UV-enhanced SiPMs in noble-gas detectors is discussed. 

%%%%%%%%%%%%%%%%%%%%%%%%%%%%%%% %%%%%%%%%%%%%%%%%%%%%%%%%%%%%%%%%%%%%%%%%%                                      Section 2                                    %%%%%%%%%%%%%%%%%%%
%%%%%%%%%%%%%%%%%%%%%%%%%% %%%%%%%%%%%%%%%%%%%%%%%%%%%%%%%%%%

\section{PDE measurement method}
\label{sec:principles}

\subsection{Principle}

The photon detection efficiency is the percentage of incident photons that are recorded by the SiPM, i.e. that are converted into a measurable electrical charge. SiPMs like other silicon-based photodetectors, have quantum efficiency (QE) close to 100\% for visible light. This is the probability that an electron-hole pair is generated by a photon incident on the silicon lattice.    
 The overall PDE is however smaller due to different factors like the micro-cells structure, leading to a reduced coverage of the active area (fill factor of 30\textendash60\%), and the probability that electrons and holes initiate an avalanche or electrical breakdown in the depleted region (P$_{av}$).  The dead time or recovery time of the micro-cells is also a factor that limits the PDE at high illumination levels. The PDE is defined by 

%%%%%%
\begin{equation}
\label{eq:pde}
 PDE = QE \times fill\_factor \times P_{av} 
\end{equation}
%%%%%%
 
The experimental determination of the PDE is provided by comparing the number of photons simultaneously incident on the SiPM (N$_{inc}$) with the number of photons recorded (N$_{rec}$) or the number of the SiPM cells excited. The relation between the number of recorded photons and the number of incident photons on the SiPM is expressed by \cite{{Hamamatsu:MPPC},{Grodzicka:2011}}
%
%%%%%%
\begin{equation}
\label{eq:Nrec_vs_Ninc}
 N_{rec} = N_{pixels} \times [1 - exp(- \frac { PDE \times N_{inc} }{N_{pixels}}) ]    
\end{equation}
%%%%
%
where N$_{pixels}$ is the total number of pixels in the SiPM. In the case of very low illumination levels with N$_{inc} \ll $  N$_{pixels}$, the response of the SiPM is linear as Eq.~(\ref{eq:Nrec_vs_Ninc}) can be approximated to 

%%%%
\begin{equation}
\label{eq:Nrec_approx}
 N_{rec} = PDE \times N_{inc}     
\end{equation}
%%%%

The number of recorded photons in the SiPM is determined from its output current (I$_{SiPM}$) and gain, which is provided by the manufacturer for a specific voltage at room temperature: 
%%%%
\begin{equation}
\label{eq:Nrec}
 N_{rec} = \frac{I_{SiPM}}{gain \times q_e} 
\end{equation}
%%%% 

\noindent q$_e$ being the elementary charge. The number of incident photons is determined using a calibrated photosensor. 
In our case a photomultiplier tube (PMT), model R8520-0SEL from Hamamatsu, was used and operated without gain.  
N$_{inc}$ is obtained from the photocurrent of the PMT collected at its first dynode (I$_{PMT}$), according to 
  
%%%%
\begin{equation}
\label{eq:Ninc}
N_{inc} = {\frac{I_{PMT} }{QE_{PMT} \times CE \times q_e} } \times {\frac{SiPM\ active\ area}{PMT\ active\ area}}
\end{equation}
%%%% 
\\

\noindent where QE$_{PMT}$ is the PMT quantum efficiency and CE its collection efficiency. This latter is optimal (100\%) when the voltage between the photocathode and the first dynode (V$_{Dy1}$) is greater than 100 V \cite{Hamamatsu:PMT}. For our measurements of the incident number of photons V$_{Dy1}$ was set to 240~V. 
 The difference between the number of photons incident on the SiPMs and on the reference PMT due to their different active areas (3~mm$\times$3~mm versus 20.5~mm$\times$20.5~mm) is taken into account in Eq.~(\ref{eq:Ninc}). 

In the method used here for the PDE measurement, known as {\it photocurrent} method \cite{Bonanno:2009}, the total number of recorded photons is not corrected for cross talk and after-pulses.
In the measurement of the PDE for applications requiring the precise determination of the absolute number of photoelectrons induced by incident photons on the SiPMs, the contribution of cross talk and afterpulses should be subtracted. 
In our present studies, the photocurrent method, also used by Hamamatsu, is adequate for our purpose of comparing the spectral response of different SiPM prototypes and for comparing our PDE measurements with those performed by Hamamatsu. 
 
%%%%%% experimental setup
\subsection{Experimental setup} \label{sec:setup}

\begin{figure}[tbhp!]
\begin{center}
\includegraphics[width=0.65\textwidth]{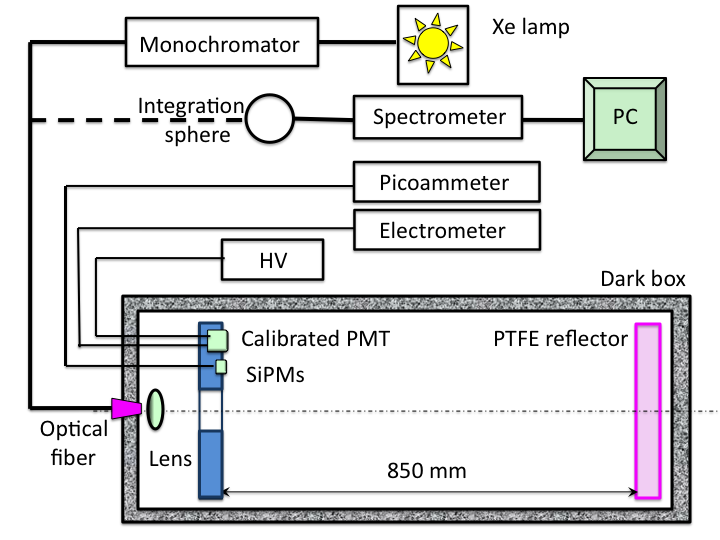} 
\end{center}
\vspace{-0.5cm}
\caption{Experimental setup used for the PDE measurements.}
\label{fig:setup} 
\end{figure}

A scheme of the experimental setup used is shown in figure~\ref{fig:setup}. It consists of a dark box of about 1~m length enclosing in one side the support plane of the photosensors (reference PMT and SiPM samples placed right beneath it) and in the opposite side a polished PTFE reflector illuminated by the light source placed behind the photosensors plane. The PTFE plane is a reflecting diffusor whose reflectance is in the range of 50\textendash70\% for the UV region and close to 98\% for visible light \cite{Silva:2009}. This diffusor is placed at a distance long enough to ensure a uniform (within 1\% in the visible region) and low level illumination of the SiPMs, typically $< 1.4\times10^8$ photons/s. 

The light source used is a xenon lamp (Hamamatsu Photonics E7536, 150 W) emitting in continuous mode and coupled to a monochromator for the selection of the input light wavelengths down to 250~nm, with a maximum standard deviation of 3~nm. The light is transported through a quartz optical fiber coupled to the dark box. A diffraction lens and various collimators were used to maintain the illumination level of the SiPMs below $1.4\times10^8$ photons/s for the PDE measurements.
 A spectrometer (Hamamatsu photonics Multichannel Analyzer C10027) was used to record the spectrogram of the input light.

%%%%%%

%%%%%%%%%%%%%%%%%%%%%%%%%%%%%%% %%%%%%%%%%%%%%%%%%%%%%%%%%%%%%%%%%%%%%%%%%                                      Section 3                                  %%%%%%%%%%%%%%%%%%%
%%%%%%%%%%%%%%%%%%%%%%%%%% %%%%%%%%%%%%%%%%%%%%%%%%%%%%%%%%%%

\section{Results and discussion} \label{sec:results}

%%%% results
The measurement of the PDE for the different SiPM prototypes was first performed using LEDs emitting at different wavelengths in the dark box setup represented in figure~\ref{fig:setup}. The LED was placed near the photosensors plane and was oriented towards the PTFE plane used as a reflecting diffusor. 
Different light intensities were obtained by varying the LED bias voltage in a range where the response of the SiPMs is linear with the number of incident photons measured by the reference PMT, as expressed in Eq.~(\ref{eq:Nrec_approx}). 
The dark current of the photosensors was first measured and subtracted from the photocurrent induced by the incident light.
The number of incident photons was drawn from the PMT's photocurrent, normalized to the SiPMs area as expressed in Eq.~(\ref{eq:Ninc}). The number of recorded photons, or number of photoelectrons produced in the SiPMs, was determined from the SiPM photocurrent as expressed in Eq.~(\ref{eq:Nrec}).

The measurements were performed at room temperature around 25$^{\circ}$C. The four SiPMs were operated at the voltages recommended by Hamamatsu for a nominal gain of $7.5\times10^{5}$. The variations in the gain induced by small changes of temperature were taken into account assuming a linear dependence of the gain with the temperature in the small range of 
($25\pm1$)$^{\circ}$C. 
The SiPM nominal gain and the ratio of photocurrents at different temperatures close to 25$^{\circ}$C were used to perform the correction of the gain for the correct determination of the PDE.  
 
\begin{figure}[tbhp!]
\begin{center}
\includegraphics[width=0.65\textwidth]{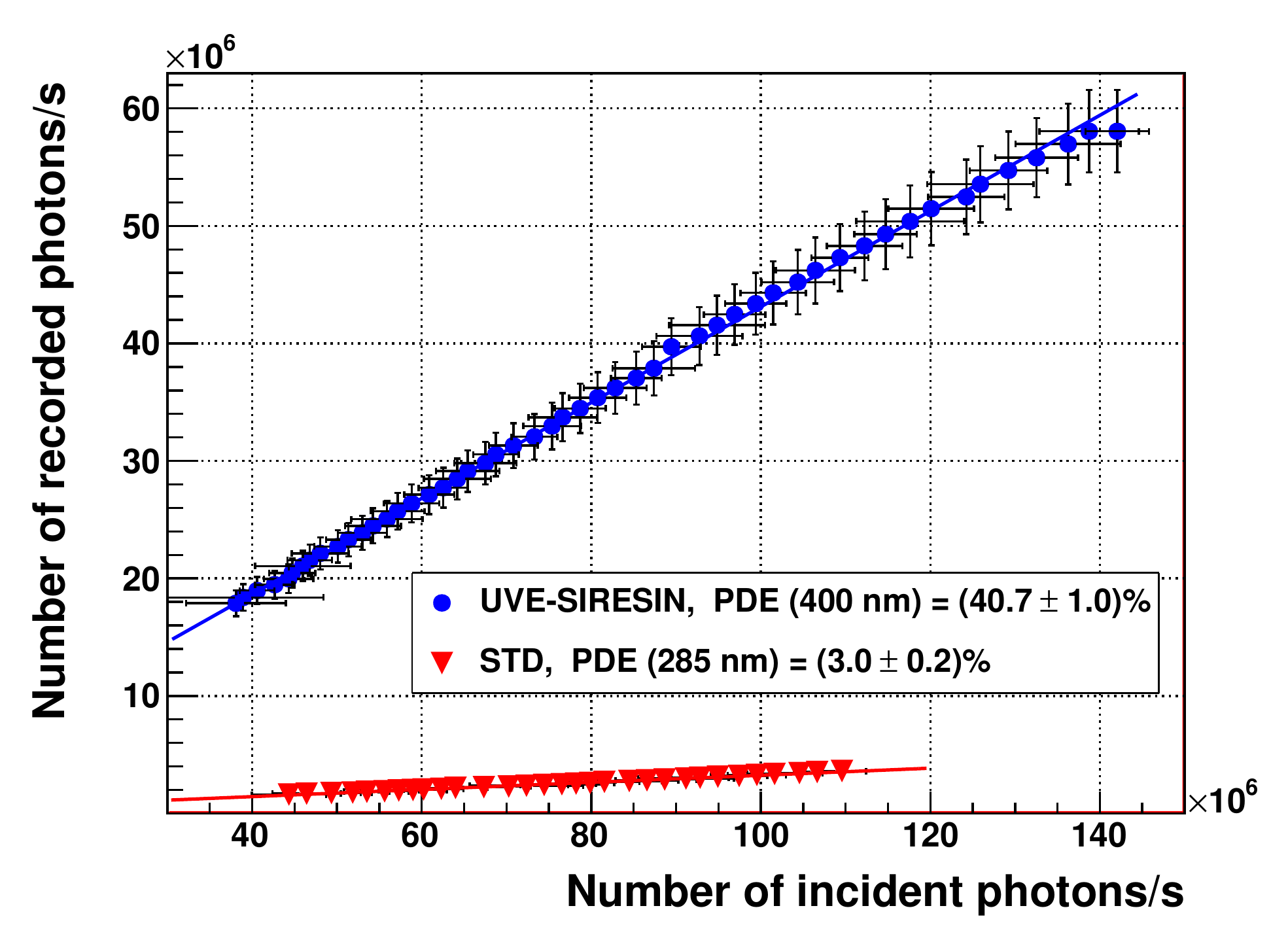} 
\end{center}
\vspace{-0.5cm}
\caption{Number of recorded photons in the UVE-SIRESIN  and STD SiPM prototypes as a function of the number of incident photons, measured respectively at 400 nm $\pm$ 5 nm and at 285 nm $\pm$ 5 nm. The PDE is determined from the linear fit to the data.}
\label{fig:Nrec_vs_Ninc} 
\end{figure}

A plot of the number of recorded photons in the UV-enhanced SiPM with silicon resin window (UVE-SIRESIN) and in the standard SiPM (STD) is shown in figure~\ref{fig:Nrec_vs_Ninc} as a function of the number of incident photons of 400 nm and 285 nm, respectively. For illumination levels below $1.4\times10^8$ photons/s, the response of the SiPMs is linear, which allows the determination of the PDE as the slope of the linear fit to the data points.
The PDE values obtained ($40.7\pm1.0$)\% at 400 nm for the UVE-SIRESIN sample and ($3.0\pm0.2$)\% at 285 nm for the STD sample) are compatible with the typical PDE values provided 
by Hamamatsu (40\% at 400 nm for the UVE-SIRESIN sample and 4\% at 285 nm for the STD sample).

The preliminary studies with LEDs at various illumination intensities allowed us to determine the illumination level required for the PDE measurements using the xenon lamp coupled to the monochromator and the spectrometer as shown in figure~\ref{fig:setup}. The light intensity from the xenon lamp was reduced using collimators after the diffraction lens in the dark box and by increasing the distance between the photosensors plane and the PTFE reflector.

%%%%%%

\begin{figure}[tbhp!]
\begin{center}
\includegraphics[width=0.65\textwidth]{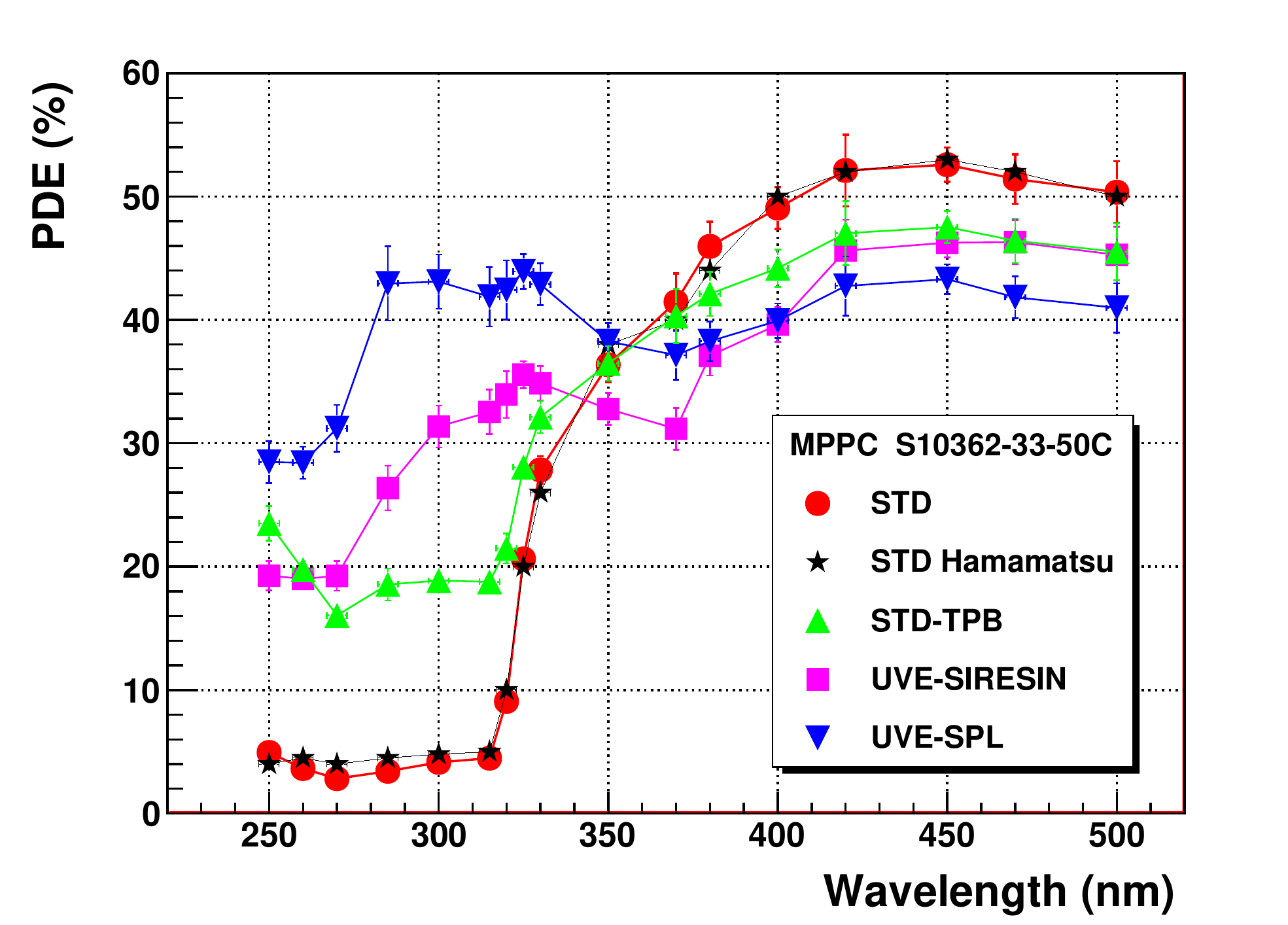} 
\end{center}
\vspace{-0.5cm}
\caption{PDE as a function of wavelength for the four SiPM prototypes considered in this study.
The typical PDE values of the standard MPPC S10362-33-50C from Hamamatsu are shown for comparison. These measurements were performed at 25$^\circ$C and include effects of cross-talk and after-pulses.}
\label{fig:pde_allsamples} 
\end{figure}

The behavior of the PDE of the standard SiPM as a function of wavelength in the range 250\textendash500 nm is in good agreement with Hamamatsu values, as shown in figure~\ref{fig:pde_allsamples}.
%
%%%%
%%%%%%
%
The PDE curves for the UV-enhanced SiPM prototypes (with silicon resin window and without window) and for the standard prototype coated with TPB are also depicted in figure~\ref{fig:pde_allsamples}. Hamamatsu measurements of the PDE for UV-enhanced SiPMs have not been published. These prototypes are not commercially available and were sent to us for research and development of UV-sensitive SiPMs, which Hamamatsu aims to be applied in UV applications and noble-gas detectors.  

As seen in figure~\ref{fig:pde_allsamples}, the standard SiPM has shown the best response in the visible region above 350 nm, compared to the other prototypes, with the maximum PDE of ($52.6\pm1.4$)\% at 450 nm. The prototype coated with TPB has shown lower PDE than the standard SiPM in the visible spectrum, as expected. This decrease of about 10\% in the PDE is related to the fraction of visible light absorbed or reflected in the TPB layer.  In the visible region, the PDE values of the UV-enhanced prototypes are about 20\% lower than that of the standard SiPM. This trend is confirmed by Hamamatsu. At 350 nm, the four SiPM samples have similar PDE which varies between 33\% and 38\%. 
At lower wavelengths, the PDE of the standard SiPM decreases considerably as already known, reaching values between 3\% and 5\% in the range 250\textendash315 nm.
In contrast, the UV-enhanced SiPM prototypes have a considerably higher PDE than the standard SiPM in this region, with  maximum values at 325 nm. 
At this wavelength, the PDE obtained for the windowless prototype is ($43.9\pm1.4$)\% and for the prototype with a silicon resin window it is ($35.6\pm1.1$)\%. 

The conversion of the UV light by the TPB-coating on the standard SiPM increases 
significantly the response of the standard photosensor at wavelengths below 350 nm. The PDE of the coated sample reaches an optimal value close to 20\% in the region 250\textendash315 nm with a slight increase up to 24\% at 250 nm. 
This trend was confirmed by repeating the measurements several times. This PDE value is half the PDE of the windowless prototype in the UV region but it is however about 6 times higher than for the non-coated SiPM.

%%%%% discussions
At 250 nm, the highest PDE value measured is ($28.5\pm1.7$)\% for the UVE windowless SiPM.
This photon detection efficiency is comparable to the quantum efficiency of standard PMTs.
The excellent response in the UV range of the UVE windowless SiPM prototype indicates it is the best choice for UV applications, especially at 325 nm where its response is optimal. The UVE sample with silicon resin window has the advantage
of having its silicon active area protected from possible damaging environmental agents.
This prototype, although with worse response in the UV region relative to the windowless prototype,
is more robust and reliable for long term operation.

In noble-gas detectors, if shifting the VUV light with an additive or dopant like trimethylamine (TMA) 
\cite{{Cureton:1981},{Nygren:2011}} is possible, the UV-enhanced SiPM with a silicon resin window would be the best choice for optical imaging. The solution of coating the standard SiPMs with TPB, adopted in experiments like NEXT, provides the possibility to increase substantially the response of the standard SiPMs in the UV region, as shown in the present work, and in the VUV (175 nm), as shown in reference \cite{Alvarez:2012-1}. TPB is however an organic molecule requiring 
stringent conditions to maintain long term UV conversion properties. The development by Hamamatsu of VUV sensitive SiPMs is therefore of utmost interest for optical imaging in noble-gas detectors.  

%% %%%%  outlook    
The experimental setup used in this work did not allow PDE measurements at lower wavelengths as this would require a vacuum box and a monochromator for tuning in the VUV light region. More measurements in the VUV range are necessary since it is not possible from the present measurements to infer the behavior of the PDE of the studied SiPM prototypes at lower wavelengths.  

%%%%%%%%%%%%%%%%%%%%%%%%%%%%%%% %%%%%%%%%%%%%%%%%%%%%%%%%%%%%%%%%%%%%%%%%%                                      Section 4                                    %%%%%%%%%%%%%%%%%%%
%%%%%%%%%%%%%%%%%%%%%%%%%% %%%%%%%%%%%%%%%%%%%%%%%%%%%%%%%%%%

\section{Conclusions}
The photon detection efficiency (PDE) of UV-enhanced SiPMs provided by Hamamatsu was determined in the spectral region 250\textendash500 nm and was compared to the PDE of the standard SiPMs of the same model (S10362-33-50C), coated and non-coated  with TPB.  
The standard SiPM has shown the best response in the visible region, as expected, with a maximum PDE close to 53\% at 450 nm.
In the UV range 250\textendash350 nm, the windowless UV-enhanced SiPM has shown the best response with a maximum PDE close to 44\% at 325 nm. In this range, the UV-enhanced SiPM with silicon resin window has a similar although lower PDE with a maximum close to 36\% at 325 nm. 
The PDE of the standard SiPM coated with 0.1~mg/cm$^2$ of TPB is typically between 20\% and 24\% in the UV region. This coated SiPM has worse performance than the UV-enhanced SiPMs, but its PDE is increased by about a factor 6 with respect to the PDE of the non-coated SiPM. 

For applications using UV light and for imaging in noble-gas detectors where the VUV scintillation from xenon or argon is shifted to the UV by an additive like TMA, the UV-enhanced SiPMs without window and with silicon resin window could be used as efficient photosensors for optical imaging.
The silicon resin window, despite of reducing the photon detection efficiency, allows the protection of the silicon layer from damaging environmental agents. It ensures therefore a more reliable and durable lifetime of the SiPMs.

More measurements in the spectral range below 250 nm are necessary to investigate the behavior of the UV-enhanced and 
TPB-coated SiPMs used in this work.   
 %%%%%%%%%%%%%%%%%%%%%%%
 
\acknowledgments
We acknowledge the Spanish MICINN for the Consolider Ingenio grants under contracts CSD2008-00037, CSD2007-00042 and for the research grant under contract FPA2009-13697-C04-01 part of which comes from FEDER funds. We also acknowledge support from the Portuguese FCT and FEDER through program COMPETE, projects PTDC/FIS/103860/2008 and PTDC/FIS/102110/2008.\\
We thank Hamamatsu Photonics for providing us free samples of UV-enhanced SiPMs and invaluable information on PDE measurements. 
We thank Henk Bolink from ICMOL for his invaluable collaboration. 

%%%%%%%%%%%%%%%%%%%%%%%


\begin{thebibliography}{9}

\bibitem{Garutti:2011} E. Garutti, \emph{Silicon photomultipliers for high energy physics detectors}, 
\jinst{6}{2011}{C10003}.

\bibitem{Ieki:2009} K. Ieki, \emph{MPPC for T2K Fine-Grained Detector}, \emph{PoS (PD09) 023.}

\bibitem{Alvarez:2012-2} NEXT collaboration:  V. Álvarez et al., \emph{NEXT-100 Technical Design Report (TDR). Executive summary}, \jinst{7}{2012}{T06001}.

\bibitem{McConkey:2011} N.~McConkey et al., \emph{Optical Readout Technology for Large Volume Liquid Argon Detectors}, {\emph{Nucl. \ Phys. (Proc. Suppl.)} {\bf B}~215~(2011)~255.}

\bibitem{Alvarez:2012-1} NEXT collaboration: V. Álvarez et al., \emph{SiPMs coated with TPB : coating protocol and characterization for NEXT}, \jinst{7}{2012}{P02010}.

\bibitem{Lightfoot:2009} P.K.~Lightfoot, G.J.~Barker, K.~Mavrokoridis, Y.A.~Ramachers and N.J.C.~Spooner, \emph{Optical readout tracking detector concept using secondary scintillation from liquid argon generated by a thick gas electron multiplier}, \jinst{4}{2009}{P04002}. 

\bibitem{Henke:data} http://henke.lbl.gov/optical$_{-}$constants.

\bibitem{LeiShi:2012} L. Shi and S. Nihtianov, \emph{Comparative study of Silicon-Based Ultraviolet Photodetectors}, {\emph{IEEE Sensors Journal}, Vol. 12, No.7, July 2012}.

\bibitem{LeiShi:2011} L. Shi, S.Nihtianov, L.K. Nanver, F. Scholze, and A. Gottlwald, \emph{High-sensitivity high-stability silicon photodiodes for DUV, VUV and EUV spectral ranges}, \emph{Proc. SPI, vol. 8145, pp.21-25, Aug. 2011}. 

\bibitem{Vanlathem:2006} E. Vanlathem et al., \emph{Novel Silicon Materials for LED Packaging and Opto-electronics devices},
\emph{Proc. SPIE, vol. 6192 (2006). doi:10.1117/12.664918}. 

\bibitem{DeGroot:2004} J.V. DeGroot Jr., A. M. Norris, S. O. Glover, T. V. Clapp, \emph{Highly transparent silicon materials}, \emph{Proc. SPIE, vol. 5517 (2004). doi:10.1117/12.557665}.

\bibitem{Hamamatsu:MPPC} \emph{MPPC technical information. http://jp.hamamatsu.com/products/division/ssd/}

\bibitem{Grodzicka:2011} M.Grodzicka et al., \emph{Effective dead time of APD cells of SiPM},
\emph{IEEE Nucl.\ Sci.\ Sym.\ Conf.\ Rec. (2011)~553}.

\bibitem{Hamamatsu:PMT}  \emph{Photomultiplier Tubes. Basics \& Applications. http://sales.hamamatsu.com}.

\bibitem{Bonanno:2009} G. Bonnano et al., \emph{Precision measurements of Photon Detection Efficiency for SiPM detectors}, { \emph{Nucl. \ Instrum. \  Meth.} {\bf A 610}~(2009)~93}.

\bibitem{Silva:2009} C.Silva et al., \emph{Reflectance of Polytetrafluorothylene (PTFE) for Xenon Scintillation Light}, \emph{J. Appl. Phys. 107 (2010) 064902. doi:10.1063/1.3318681}.

\bibitem{Cureton:1981} C.G.Cureton et al., \emph{The photonics of tertiary Aliphatic Amines},\emph{Chem. Phys. 63 (1981)~31}.

\bibitem{Nygren:2011} David Nygren. \emph{Can the ``intrinsic'' energy resolution in xenon be surpassed ?},
\emph{J. Phys.: Conf. Ser. {\bf 309} (2011) 012006.  doi:10.1088/1742-6596/309/1/012006.}

\end{thebibliography}
\end{document}